# Watching outside while under a carpet cloak of invisibility

Jin-Zhu Zhao, De-Lin Wang, Ru-Wen Peng*, and Mu Wang

National Laboratory of Solid State Microstructures and Department of Physics, Nanjing University, Nanjing 210093, China

**Abstract**

We demonstrate in this letter a unique approach for watching outside while hiding in a carpet cloaking based on transformation optics. Unlike conventional carpet cloaking, which screens all the incident electromagnetic waves, we break the cloak and allow incident light get into the carpet. Hence outside information is detected inside the cloak. To recover the invisible cloaking, complementary techniques are applied in the broken space. Consequently, a hiding-inside-and-watching-outside (HIWO) carpet cloak is sewed, which works as a perfectly invisible cloaking and allows surveillance of the outside at the same time. Our work provides a strategy for ideal cloak with "hiding" and "watching" functions simultaneously.

* To whom correspondence should be addressed. Electronic address: rwpeng@nju.edu.cn

Invisible cloak has been conceived by mankind for a long time. Very recently this imagination has turned to be possible. Pendry *et al*. [1] proposed a scheme to design a cloaking of objects from electromagnetic fields by using transformation optics [2]. Leonhardt [3] developed optical conformal mapping for an invisibility device. Inspired by the theoretical strategies, metamaterial microwave cloaking has been experimentally realized for the first time [4]. However, some problems remain challenging, such as singular parameter and narrow-band limit of the cloak [1]. In order to solve parameter singularity of the cloak, carpet cloaking has been proposed to give all objects the appearance of a flat conducting sheet [5], which has been experimentally achieved at microwave [6] and optical [7] frequencies, respectively. The early approaches tended to hide an object inside the cloaked domain. Recently Lai *et al* proposed a different approach which allows to cloak the object at a distance outside the cloaking shell based on complementary media [8]. Up to now, the concept of cloaking has been extended from electromagnetic wave to both acoustic wave [9] and matter wave [10] on the analogy of wave equations.

However, current invisibility cloaks are still far from satisfactory. People always expect that a man in invisibility cloak can not only hide, but also be able to watch the outside. A perfect cloaking should be undetectable by electromagnetic (EM) detection because it is equivalent to a curved but empty EM space created by coordinate transformation. Very recently Zhang and Wu [11] proposed a method to detect this curved EM space by shooting a fast-moving charge particle through it. With this approach, people outside the cloak detect the cloak instead of the hidden people detecting the outside world. It should be emphasized that the people can indeed carry out detection based on the concept of illusion optics, where people see through a wall by opening a virtual hole on it [12]. However, by doing so, people in front of the wall and behind the wall can "see" each other, *i.e.*, the device does not function as an invisible cloak. In the complementary media cloak proposed in Ref. 8, the cloak is invisible yet the object is exposed to the external field and can certainly "watch" outside. Recently, Alù and Engheta proposed a concept of clocking a sensor [13], which has opened a way of sensing with light through a nearly invisible detector [14]. Thereafter, the concept of cloaked sensing has been extensively studied [15,16].

In this letter, we propose a scheme for watching the outside while hiding under a carpet cloaking based on transformation optics. Unlike conventional carpet cloaking, which screens all the incident electromagnetic waves, we break the cloak and allow incident light get into the carpet. Hence outside information is detected inside the cloak. To recover the invisible cloaking, a complementary medium is embedded in the broken space. Meanwhile, in order to detect incident light, an ordinary sensor (OS) and an "anti-sensor" (AS) are designed in the broken space. The OS detects the incoming signal yet breaks the cloaking again, whereas the AS compensates the breaking induced by the OS and recovers the invisible carpet. In this way, our hiding-inside-and-watching-outside (HIWO) carpet cloak functions not only as a perfect invisible cloaking, but as the surveillance under the cloaking as well.

The concept of complementary media is regarded as a special type of transformation media [17-20], which achieves the potential applications in perfect lens [17], novel imaging devices [8], superscatterers [21], cylindrical superlens [22], and anticloak [23]. Very recently, complementary media have been successfully applied to create a cloak outside the cloaking shell [8], and to design illusion device [12]. Here we apply complementary media to achieve HIWO carpet cloak. According to transformation optics [24], when a space is transformed into a different shape, the permittivity and the permeability in two spaces are connected by

$$\varepsilon^{i'j'} = \frac{1}{\det(\Lambda_i^{i'})} \Lambda_i^{i'} \Lambda_j^{j'} \varepsilon^{ij}, \quad (1)$$

$$\mu^{i'j'} = \frac{1}{\det(\Lambda_i^{i'})} \Lambda_i^{i'} \Lambda_j^{j'} \mu^{ij}, \quad (2)$$

respectively. Here $\varepsilon^{ij}$ and $\mu^{ij}$ are the permittivity and the permeability tensors in original space $(i, j = 1, 2, 3)$, and $\varepsilon^{i'j'}$ and $\mu^{i'j'}$ are the permittivity and the permeability tensors in the new space, and $\Lambda_i^{i'} = \frac{\partial x^{i'}}{\partial x^i}$ is the Jaccobian tensor of the transformation. Then it is possible to design the complementary media by operating a specific coordinate transformation to fold a piece of space into another. One of the typical examples of complementary media is perfect lens. As shown in Fig. 1(a), a perfect lens [17] can be a slab designed by the coordinate transformation of $y' = -y$ for $-L < y < 0$, which means folding an air slab of $-L < y < 0$ into the slab of $0 < y' < L$ with $\varepsilon' = -1$ and $\mu' = -1$. (Here L is the width of the slab.) Due to the cancellation of optical path, the region (-L, L) does not exist optically. Now we consider an object with permittivity $\varepsilon_0$ and permeability $\mu_0$ located in the air (Fig. 1(b)). We can design a slab of complementary media by the coordinate transformation of $y' = -y$ for $-L < y < 0$, which results in a slab of $\varepsilon' = -1$ and $\mu' = -1$. A complementary medium with permittivity $-\varepsilon_0$ and permeability $-\mu_0$ is embedded inside the slab. Similar to the perfect lens, the object and the complementary media are optically cancelled each other.

In order to watch outside under the original carpet cloaking [5], we have to break the carpet cloaking and let some of incident electromagnetic fields get into the cloaking, which eventually makes the object under the carpet visible. However, if we separate the broken space into two parts: Part I filled by a complementary medium with permittivity $\boldsymbol{\varepsilon}'$ and permeability $\boldsymbol{\mu}'$ and Part II filled with air (Fig. 1(c)), the complementary medium (part I) and air in Part II are optically cancelled each other. In this way, the invisible carpet is recovered. As illustrated in Fig. 1(c), the complementary medium can be obtained by the coordinate transformation of folding and compressing the volume of air ($c < y < b$) into the layer of complementary medium ($b < y' < a$). Here a, b, and c denote the coordinates of the interfaces of different media as shown in Fig. 1(c). If we consider a coordinate transformation $y' = f(y)$, where $f(y)$ is a continuous function satisfying $f(c) = a$ and $f(b) = b$. Parameters $\boldsymbol{\varepsilon}'$ and $\boldsymbol{\mu}'$ in the complementary layer can be obtained based on Eqs. (1) and (2). For simplification, we take linear relation $y' = f(y) = \frac{b-a}{b-c} y + \frac{b(a-c)}{b-c}$. It follows that $\boldsymbol{\varepsilon}'$ and $\boldsymbol{\mu}'$ are expressed as

$$\boldsymbol{\varepsilon}' = \boldsymbol{\mu}' = \begin{pmatrix} \frac{b-c}{b-a} & 0 & 0 \\ 0 & \frac{b-a}{b-c} & 0 \\ 0 & 0 & \frac{b-c}{b-a} \end{pmatrix}. \qquad (3)$$

Consequently Part I and Part II are optically cancelled, and the reflection from the breaking on the carpet is the same as that from the original invisible carpet.

Now that incident electromagnetic fields come into the broken space in our device, we can put an ordinary sensor (OS) with permittivity $\varepsilon_0$ and permeability $\mu_0$ inside Part II to detect the incident wave, as schematically shown in Fig. 1(d). However, introducing OS makes the cloak visible. To avoid this penalty, based on transformation optics, we introduce an "anti-sensor" (AS) embedding in the complementary layer (Part I, as shown in Fig. 1(d)) with parameters

$$\begin{cases} \boldsymbol{\varepsilon'}_0 = \varepsilon_0 \boldsymbol{\varepsilon'} \\ \boldsymbol{\mu'}_0 = \mu_0 \boldsymbol{\mu'} \end{cases}. \qquad (4)$$

In this way, OS and AS are optically cancelled. The object under the carpet becomes invisible, yet the OS can still detect the incident wave from the invisible carpet. This unique design allows watching outside by hiding a sensor under the carpet cloaking without being detected from the outside.

To verify this unique idea, we carry out full-wave simulation with commercial software (COMSOL Multiphysics 3.5 solver). The incident wave is taken as a Gaussian beam at a wavelength of 0.3$m$, which is transverse magnetic (TM) polarization (H along the z direction). (Here intensity distribution of the beam follows Gaussian function in space.) The incident light approaches the cloak at 45$^o$ with respect to the ground plane. Firstly, we start from the original carpet cloak covering the bump region as shown in Fig. 2(a). Reflection at 45$^o$ is clearly observed, and the distribution of magnetic fields outside the cloak is the same as that from a flat plate. In this case, the incident wave is totally reflected by the highly reflective boundary, and the electromagnetic wave cannot get into the bump. Consequently the incident wave cannot be detected from the inside of the bump. Secondly, we break the part of highly reflective boundary. As shown in Fig. 2(b), the electromagnetic wave propagates into the bump region, and the reflected beam is scattered into two beams. The magnetic field outside the cloak becomes quite different from that from a flat plate. Obviously, the cloak is broken and the bump becomes visible. Thirdly, we recover the invisible cloak with complementary media. The broken space is amended in the following way: Part I (b<y<a) is filled by a complementary medium with permittivity **ε'** and permeability **μ'**, and Part II (b<y<c) is filled with the air (as schematically shown in Fig.1(c)). By taking *a*=0.16*m*, *b*=0.128*m*, and *c*=0, we apply linear transformation $y' = -\frac{1}{4}y + \frac{4}{25}$. Thereafter, the parameters **ε'** and **μ'** for the complementary medium are given by Eq. (3). As shown in Fig. 2(c), the magnetic field outside the carpet cloak becomes the same as that in Fig. 2(a), indicating that the invisibility cloaking is recovered. Moreover, some incident waves get into the bump region, which makes it possible to detect the external information inside the cloak. We can push one step further to detect the incident wave inside the cloak and finally obtain a HIWO carpet cloak. For example, we put an ordinary sensor with the parameters $\varepsilon_0=2$ and $\mu_0=2$ in Part II and an anti-sensor with the parameters of $\boldsymbol{\varepsilon}'_0$ and $\boldsymbol{\mu}'_0$ in Part I (Fig. 1(d)). Here $\boldsymbol{\varepsilon}'_0$ and $\boldsymbol{\mu}'_0$ are given by Eqs. (3) and (4). As shown in Fig. 2(d), the magnetic field distribution outside the carpet cloak keeps the same as that in Fig. 2(a), indicating that the invisibility cloaking resumes. Inside of the cloak, as shown in Fig. 2(d), the filed distribution indicates that the incident light enters the cloak already, which is different from that of Fig. 2(a).

In order to confirm that incident wave can indeed be detected in the HIWO carpet cloak, we show the distribution of magnetic field in the sensor when the incident waves with different wavelengths shine on the identical HIWO carpet cloak, as shown in Fig. 3(a)-(h). When the wavelength of incident wave varies, the magnetic field distributions outside the cloaks (as shown in Fig. 3(a)-(d)) are the same as those of a flat plate, indicating that the cloak functions and the invisibility persists at all these wavelengths. Meanwhile, the signal detected by the sensor under the cloak is different when the wavelength of the incident wave is changed (as shown in Figs. 3(e)-(h)). Figure 3(i) shows the magnetic field along the line of $y=0.05m$, where the gray region corresponds to the position of the sensor. The intensity of the signal received by the sensor varies obviously as the wavelength of incident light is increased. These data indicate clearly that the information carried by incident wave can be detected and watching outside from HIWO carpet cloak is indeed applicable.

The above discussed HIWO cloaking consists of the original carpet cloak, the complementary medium, a sensor and an anti-sensor. It is known that original carpet cloaks have been experimentally realized at microwave frequencies [6] and at optical frequencies [7], respectively. The complementary medium, the sensor and the anti-sensor in the HIWO cloaking can be readily realized by a metamaterial with both negative permittivity and negative permeability in microwave frequency [25]. For optical frequency, three-dimensional negative refractive index material has been achieved very recently [26]. By setting $a=0.32\mu m$, $b=0.256\mu m$, and $c=0$ in a HIWO invisible cloaking and applying a metallic sensor with $\varepsilon_0=-2$ and $\mu_0=1$, and the anti-sensor with $\boldsymbol{\varepsilon'_0}$ and $\boldsymbol{\mu'_0}$ following Eqs. (3) and (4), a HIWO invisible cloaking can work at the optical wavelength of $600nm$. As illustrated in Fig. 4(a), the cloak is broken without a complementary layer and an anti-sensor. However, once we introduce a complementary layer and an anti-sensor in the device, invisibility resumes while the sensor under the cloak detects the incidence wave (as illustrated in Fig. 4(b)). Therefore, it is experimentally possible to realize the HIWO carpet cloak at microwave and optical frequencies, respectively.

In summary, we propose here a HIWO carpet cloak based on transformation optics, and demonstrate that the HIWO carpet cloak works not only as a perfectly invisible cloaking, but as surveillance under the cloaking as well. The experimental possibility to realize the design at microwave and optical regimes has been discussed. With such a unique HIWO carpet cloak, we can watch outside while hide under a carpet cloaking in the frame of transformation optics, which contributes a strategy to achieve ideal invisible cloaking with simultaneous "hiding" and "watching".

## Acknowledgments

This work has been supported by the National Natural Science Foundation of China (Grant Nos. 10625417, 50972057, and 10874068), the State Key Program for Basic Research from MOST of China (Grant Nos. 2010CB630705 and 2006CB921804), and partly by Jiangsu Province (Grant No. BK2008012).

**References:**

1. J. B. Pendry, D. Schurig, and D. R. Smith, Science **312**, 1780 (2006).

2. A. J. Ward, and J. B. Pendry, J. Mod. Opt. **43**, 773 (1996).

3. U. Leonhardt, Science **312**, 1777 (2006).

4. D. Schurig *et al.*, Science **314**, 977 (2006).

5. J. Li, and J. B. Pendry, Phys. Rev. Lett. **101**, 203901 (2008).

6. R. Liu, C. Ji, J. J. Mock, J. Y. Chin, T. J. Cui, and D. R. Smith, Science **323**, 366 (2009).

7. J. Valentine, J. Li, T. Zentgraf, G. Bartal, and Xiang Zhang, Nature Materials **8**, 568 (2009).

8. Y. Lai, H. Chen, Z. Q. Zhang, and C. T. Chan, Phys. Rev. Lett. **102**, 093901 (2009).

9. S. A. Cummer, and D. Schurig, New Journal of Physics **9**, 45 (2007).

10. Shuang Zhang, D. A. Genov, Cheng Sun, and Xiang Zhang, Phys. Rev. Lett. **100**,123002 (2008).

11. B. Zhang, and B. I. Wu, Phys. Rev. Lett. 103, 093901 (2009).

12. Y. Lai, Jack Ng, H. Y. Chen, D. Z. Han, J. J. Xiao, Z. Q. Zhang, and C. T. Chan, Phys. Rev. Lett. **102**, 253902 (2009).

13. A. Alu, and N. Engheta, Phys. Rev. Lett. 102, 233901 (2009).

14. E. J. Garcia de Abajo, Physics 2, 47 (2009).

15. Z. Ruan, and S. Fan, J. Phys. Chem. C 114, 7324 (2010).

16. A. Greenleaf, Y. Kurylev, M. Lassas, and G. Uhlmann, preprint arXiv: 0912.1872v1.

17. J. B. Pendry, Phys. Rev. Lett. 85, 3966 (2000).

18. J. B. Pendry, and S. A. Ramakrishna, J. Phys. Condens. Matter 14, 8463 (2002); 15, 6345 (2003).

19. K. Kobayashi, J. Phys. Condens. Matter 18, 3703 (2006).

20. W. Yan, M. Yan, and M. Qiu, arXiv:0806.3231.

21. T. Yang *et al.*, Opt. Express 16, 18545 (2008).

22. M. Yan, W. Yan, and M. Qiu, Phys. Rev. B 78, 125113 (2008).

23. H. Y. Chen *et al.*, Opt. Express 16, 14603 (2008).

24. D. Schurig, J. B. Pendry, and D. R. Smith, Opt. Express **14**, No. 21 9794 (2006).

25. D. R. Smith, J. B. Pendry, and M. C. K. Wiltshire, Science **305**, 788 (2004).

26. J. Valentine, S. Zhang, T. Zentgraf, E. Ulin-Avila, D. A. Genov, G. Bartal, and Xiang Zhang, Nature **455**, 376 (2008).

**Figure Captions:**

**Fig. 1 (a)** The schematics to show the complementary media (black region), $0<y'<L$, which is used to cancel the volume of air, $-L<y<0$. **(b)** The schematics to show a complementary media with an embedded complementary "image" of $-\varepsilon_0$ and $-\mu_0$, which cancels an object in the air with $\varepsilon_0$ and $\mu_0$. **(c)** The schematics to show an amended broken carpet cloaking device. The broken space is separated into two parts: Part I (b<y<a, black region) is composed of complementary medium with $\boldsymbol{\varepsilon'}$ and $\boldsymbol{\mu'}$, and Part II (b<y<c, white region) is filled with air. The red line is the boundary with perfectly reflective conductor. **(d)** The schematics of the HIWO carpet cloaking device. On the device shown in Fig. 1(c), a sensor (blue region) with permittivity $\varepsilon_0$ and permeability $\mu_0$ is embedded inside Part II (air), and an "anti-sensor" (yellow region) with $\boldsymbol{\varepsilon'_0}$ and $\boldsymbol{\mu'_0}$ is embedded inside Part I.

**Fig. 2** The plots to show the magnetic field distribution when the incident transverse-magnetic wave launches towards the ground plane at $45°$ from the left. The spacial width of the beam is $1.3m$ at the wavelength of $0.3m$, the unit for the positions (X and Y) is meters, and the cloak is within the rectangle in black line in each case. **(a)** The field distribution around an original carpet cloak as reported in R. **(b)** The field distribution around a broken carpet cloak without complementary media. **(c)** The field distribution around a broken carpet cloak with complementary medium as shown in Fig. 1(c). Here $a=0.16m$, $b=0.128m$, and $c=0$. Both $\boldsymbol{\varepsilon'}$ and $\boldsymbol{\mu'}$ are given by Eq. (3). **(d)** The field distribution around the HIWO carpet cloaking as shown in Fig. 1(d). In this devise, a sensor with $\varepsilon_0=2$ and $\mu_0=2$ is inside Part II, and an anti-sensor with $\boldsymbol{\varepsilon}'_0$ and $\boldsymbol{\mu}'_0$ is inside Part I, where $\boldsymbol{\varepsilon}'_0$ and $\boldsymbol{\mu}'_0$ are given by Eqs. (3) and (4).

**Fig. 3 (a)-(d)** The distributions of magnetic field around the HIWO carpet cloak as shown in Fig. 2(d) with different wavelength ($\lambda$) launching at $45°$ towards the ground plane from the left. The spacial width of the beam is $1.3m$ in each case. **(a)** $\lambda = 0.3m$, **(b)** $\lambda = 0.4m$, **(c)** $\lambda = 0.5m$, and **(d)** $\lambda = 0.6m$. In order to show how the sensor inside of the cloak works in each case, the detailed distributions of the magnetic fields close to the sensor (located at the triangle site) are given in **(e)-(h)**, respectively. **(i)** The intensity of magnetic field along $y = 0.05m$ for different wavelengths, the shadowed region corresponds to place where the sensor is located. One may clearly see that the signal received by the sensor varies at different incident wavelength.

**Fig. 4** The plots to show the distribution of magnetic field at optical frequency. The incident transverse-magnetic wave launches at $45°$ towards the ground plane from the left. The parameters are set as $a = 0.32\mu m$, $b = 0.256\mu m$, and $c=0$. The spacial width of the beam is $2\mu m$ at the wavelength of $\lambda = 600nm$, the unit for the positions (X and Y) is microns, and the cloak is within the rectangle in black line. **(a)** The field distribution around a broken carpet cloak. In the broken space, a sensor with $\varepsilon_0=-2$ and $\mu_0 =1$ is embedded, but no complementary layer and anti-sensor are introduced. **(b)** The field distribution around the HIWO carpet cloaking as shown in Fig. 1(d) at $\lambda = 600nm$. The broken space is separated into two parts: Part I is composed of a complementary medium with parameters $\boldsymbol{\varepsilon}'$ and $\boldsymbol{\mu}'$, and Part II is filled with the air, where $\boldsymbol{\varepsilon}'$ and $\boldsymbol{\mu}'$ are given by Eq. (3). Then a sensor with $\varepsilon_0 = -2$ and $\mu_0 = 1$ is put inside Part II, and an anti-sensor with $\boldsymbol{\varepsilon}'_0$ and $\boldsymbol{\mu}'_0$ inside Part I, respectively, where $\boldsymbol{\varepsilon}'_0$ and $\boldsymbol{\mu}'_0$ are given by Eqs. (3) and (4).

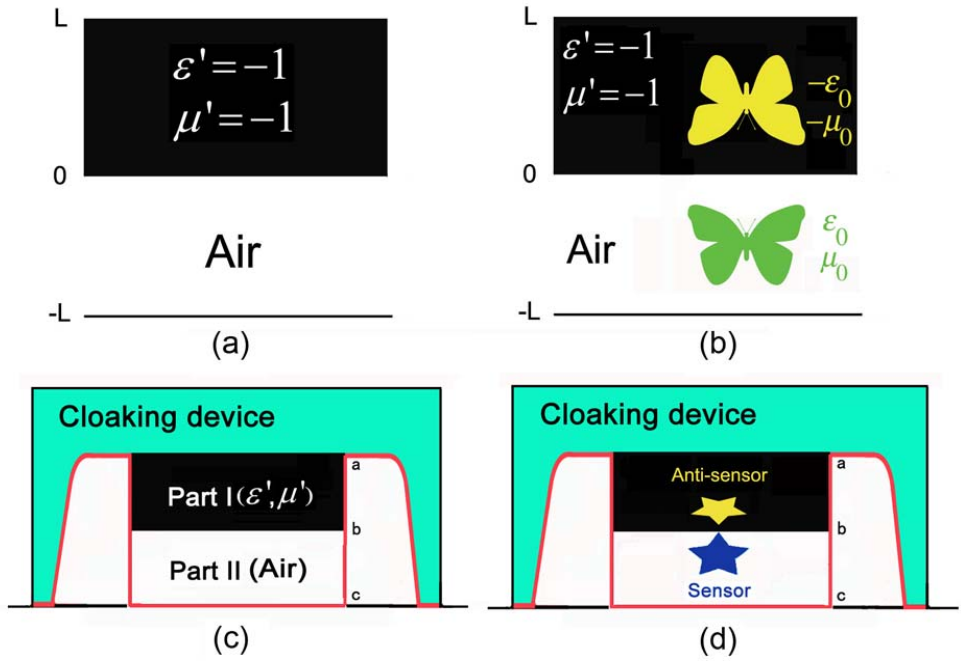

**Figure 1**

**(a)** The schematics to show the complementary media (black region), $0 < y' < L$, which is used to cancel the volume of air, $-L < y < 0$.

**(b)** The schematics to show a complementary media with an embedded complementary "image" of $-\varepsilon_0$ and $-\mu_0$, which cancels an object in the air with $\varepsilon_0$ and $\mu_0$.

**(c)** The schematics to show an amended broken carpet cloaking device. The broken space is separated into two parts: Part I (b<y<a, black region) is composed of complementary medium with $\varepsilon'$ and $\mu'$, and Part II (b<y<c, white region) is filled with air. The red line is the boundary with perfectly reflective conductor.

**(d)** The schematics of the HIWO carpet cloaking device. On the device shown in Fig. 1(c), a sensor (blue region) with permittivity $\varepsilon_0$ and permeability $\mu_0$ is embedded inside Part II (air), and an "anti-sensor" (yellow region) with $\varepsilon'_0$ and $\mu'_0$ is embedded inside Part I.

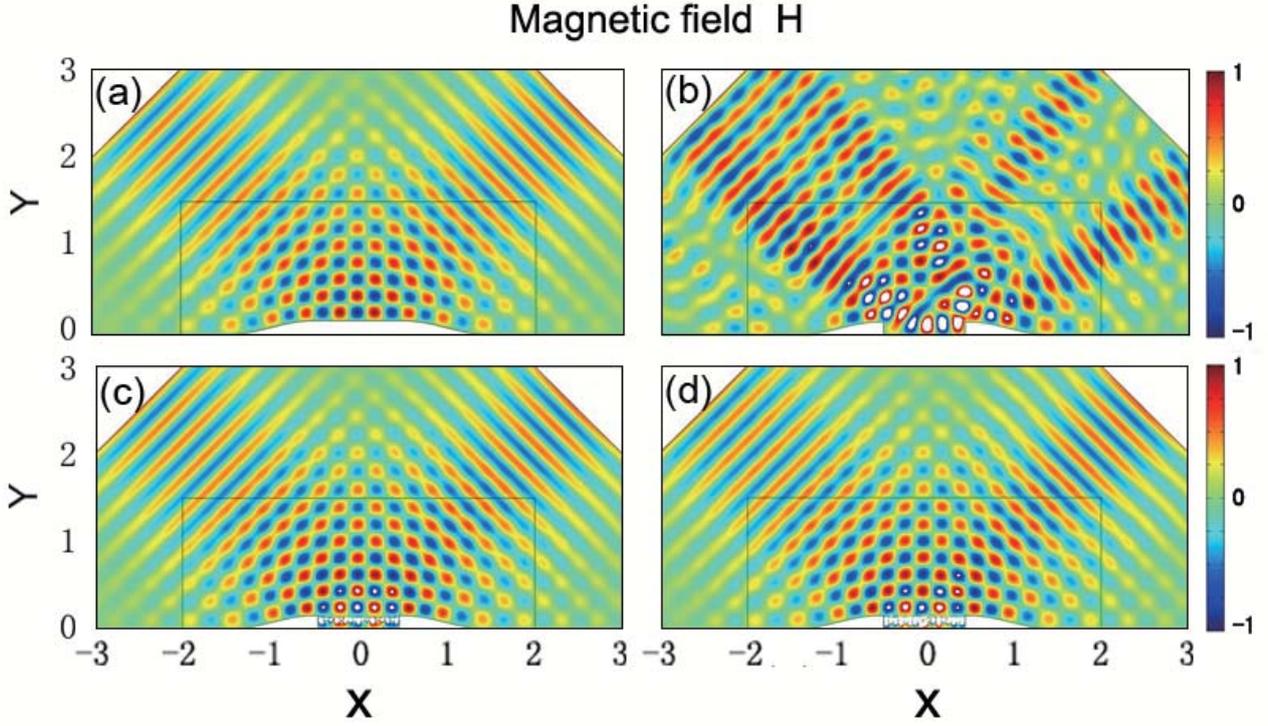

**Figure 2**

The plots to show the magnetic field distribution when the incident transverse-magnetic wave launches towards the ground plane at 45° from the left. The spacial width of the beam is 1.3$m$ at the wavelength of 0.3$m$, the unit for the positions (X and Y) is meters, and the cloak is within the rectangle in black line in each case.

**(a)** The field distribution around an original carpet cloak as reported in Ref.5.

**(b)** The field distribution around a broken carpet cloak without complementary media.

**(c)** The field distribution around a broken carpet cloak with complementary medium as shown in Fig. 1(c). Here $a=0.16m$, $b=0.128m$, and $c=0$. Both $\varepsilon'$ and $\mu'$ are given by Eq. (3).

**(d)** The field distribution around the HIWO carpet cloaking as shown in Fig. 1(d). In this devise, a sensor with $\varepsilon_0=2$ and $\mu_0=2$ is inside Part II, and an anti-sensor with $\varepsilon'_0$ and $\mu'_0$ is inside Part I, where $\varepsilon'_0$ and $\mu'_0$ are given by Eqs. (3) and (4).

**Figure 3**

**(a)-(d)** The distributions of magnetic field around the HIWO carpet cloak as shown in Fig. 2(d) with different wavelength ($\lambda$) launching at $45°$ towards the ground plane from the left. The spacial width of the beam is $1.3m$ in each case. **(a)** $\lambda = 0.3m$, **(b)** $\lambda = 0.4m$, **(c)** $\lambda = 0.5m$, and **(d)** $\lambda = 0.6m$. In order to show how the sensor inside of the cloak works in each case, the detailed distributions of the magnetic fields close to the sensor (located at the triangle site) are given in **(e)-(h)**, respectively.

**(i)** The intensity of magnetic field along $y = 0.05m$ for different wavelengths, the shadowed region corresponds to place where the sensor is located. One may clearly see that the signal received by the sensor varies at different incident wavelength.

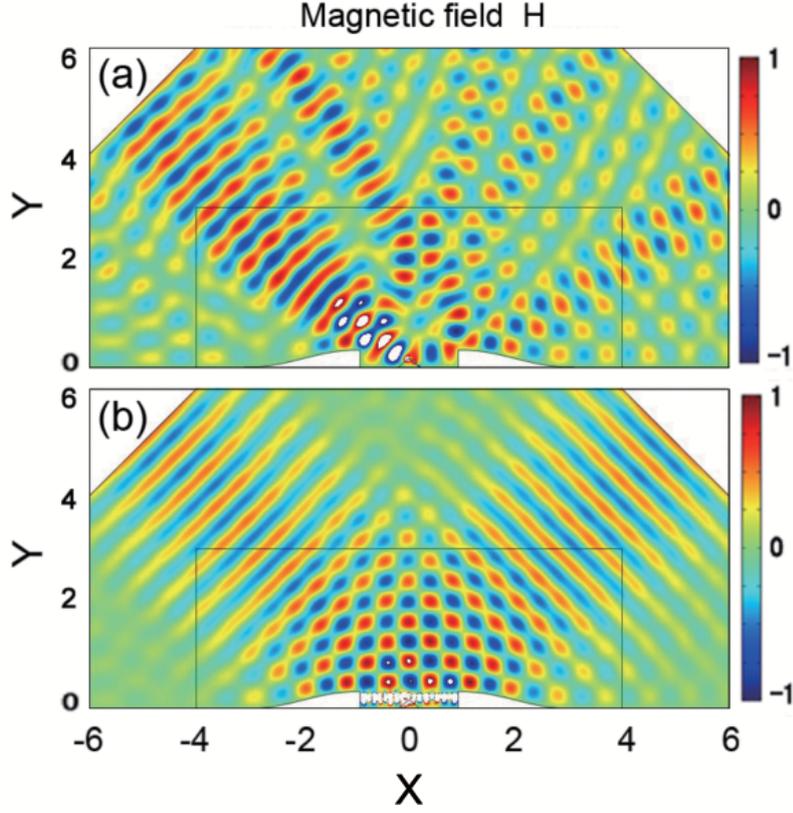

**Figure 4**

The plots to show the distribution of magnetic field at optical frequency. The incident transverse-magnetic wave launches at $45°$ towards the ground plane from the left. The parameters are set as $a = 0.32 \mu m$, $b = 0.256 \mu m$, and $c=0$. The spacial width of the beam is $2 \mu m$ at the wavelength of $\lambda = 600 nm$, the unit for the positions (X and Y) is microns, and the cloak is within the rectangle in black line.

**(a)** The field distribution around a broken carpet cloak. In the broken space, a sensor with $\varepsilon_0 = -2$ and $\mu_0 = 1$ is embedded, but no complementary layer and anti-sensor are introduced.

**(b)** The field distribution around the HIWO carpet cloaking as shown in Fig. 1(d) at $\lambda = 600 nm$. The broken space is separated into two parts: Part I is composed of a complementary medium with parameters $\boldsymbol{\varepsilon'}$ and $\boldsymbol{\mu'}$, and Part II is filled with the air, where $\boldsymbol{\varepsilon'}$ and $\boldsymbol{\mu'}$ are given by Eq. (3). Then a sensor with $\varepsilon_0 = -2$ and $\mu_0 = 1$ is put inside Part II, and an anti-sensor with $\boldsymbol{\varepsilon'_0}$ and $\boldsymbol{\mu'_0}$ inside Part I, respectively, where $\boldsymbol{\varepsilon'_0}$ and $\boldsymbol{\mu'_0}$ are given by Eqs. (3) and (4).